\let\orilabel\label 
\documentclass[pra,aps,showpacs,superscriptaddress,twocolumn,10pt]{revtex4-2}
\let\label\orilabel

\usepackage{physics}
\usepackage[cm]{fullpage}
\usepackage{graphicx}	
\usepackage[english]{babel}
\usepackage{amsmath}
\usepackage{amssymb,bm}
\usepackage{natbib}
\usepackage{xcolor}
\usepackage{bbold}
\usepackage{mathtools}
\usepackage{hyperref}

\begin{document}
\author{Erik Weiss}
\affiliation{Institut f\"ur Theoretische Physik, Universit\"at T\"{u}bingen, 72076 T\"ubingen, Germany}
\author{Marcel Cech}
\affiliation{Institut f\"ur Theoretische Physik, Universit\"at T\"{u}bingen, 72076 T\"ubingen, Germany}
\author{Stanislaw Soltan}
\affiliation{Institut f\"ur Theoretische Physik, Universit\"at T\"{u}bingen, 72076 T\"ubingen, Germany}
\author{Martin Koppenhöfer}
\affiliation{Fraunhofer Institute for Applied Solid State Physics IAF, Tullastr. 72, Freiburg 79108, Germany}
\author{Michael Krebsbach}
\affiliation{Fraunhofer Institute for Applied Solid State Physics IAF, Tullastr. 72, Freiburg 79108, Germany}
\author{Thomas Wellens}
\affiliation{Fraunhofer Institute for Applied Solid State Physics IAF, Tullastr. 72, Freiburg 79108, Germany}
\author{Daniel Braun}
\affiliation{Institut f\"ur Theoretische Physik, Universit\"at T\"{u}bingen, 72076 T\"ubingen, Germany}

\title{Pattern-based quantum functional testing}

\begin{abstract}
	With the growing number of qubits of quantum information processing devices, the task of fully characterizing these processors becomes increasingly unfeasible.  From a practical perspective, one wants to find possible errors in the functioning of the device as quickly as possible, or otherwise establish its correct functioning with high confidence.  In response to these challenges, we propose a pattern-based approach inspired by classical memory testing algorithms to evaluate the functionality of a quantum memory, based on plausible failure mechanisms.  We demonstrate the method's capability to extract pattern dependencies of important qubit characteristics, such as $T_1$ and $T_2$ times, and to identify and analyze interactions between adjacent qubits. Additionally, our approach enables the detection of different types of crosstalk effects and of signatures indicating non-Markovian dynamics in individual qubits.
\end{abstract}

  \title{Pattern-based quantum functional testing}
  \date{\today}
  \maketitle

\section{Introduction}
  Traditionally the characterization of quantum devices has focused on
  complete tomography of produced quantum states or even quantum
  channels
  \cite{chuang_prescription_1997,PhysRevA.63.020101,PhysRevLett.86.4195,PhysRevLett.90.193601,PhysRevA.64.012314,leonhardt_tomographic_1995,kosut_quantum_2008,christandl_reliable_2012,blume-kohout_robust_2012,rey-de-castro_time-resolved_2013,ohliger_efficient_2013}. Powerful
  methods such as randomized benchmarking \cite{Emerson05.2,knill_randomized_2008,bendersky_selective_2009,cramer_efficient_2010,magesan_scalable_2011,magesan_efficient_2012,gaebler_randomized_2012,onorati_randomized_2018,cross_validating_2018,nielsen_gate_2021},
  quantum process tomography\cite{chuang_prescription_1997, Poyatos1997,Nielsen00}, gate-set tomography \cite{nielsen_gate_2021,Merkel2013, Blume-kohout2017,Greenbaum2015}, 
  compressed sensing \cite{gross_quantum_2010,flammia_quantum_2012}, or extrapolation of
  error models have been introduced \cite{PhysRevLett.119.180509}.
  Bayesian experimental design and metrology 
  \cite{huszar_adaptive_2012,huszar_adaptive_2012,ferrie_self-guided_2014,kueng_near-optimal_2015,granade_practical_2016,PhysRevA.108.022602} 
  and machine learning techniques were applied
  \cite{PRXQuantum.2.020303,quek_adaptive_2021,gebhart_learning_2023}, and representations of states via
  ``classical shadows'' \cite{huang_predicting_2020} developed.  While
  these techniques have led  
  to important insights, it is clear that complete tomography of quantum
  channels becomes unfeasible for more than a few qubits due to the
  exponential growth of Hilbert space.  That problem, however, is shared
  by classical information processing devices, where, e.g.,~the number of
  possible states of a memory chip scales as $2^{N_\text{bit}}$, with the
  number of bits $N_\text{bit}$. Yet, these devices function remarkably
  well due to functional tests adapted to the hardware of the chips.
  From the perspective of a test engineer, the task is a
  decision problem: decide as quickly as possible whether the device
  under test does either \textit{not} function correctly or functions
  correctly with a high level of confidence.   With state-of-the-art quantum
  processors containing already on the order of thousand qubits, it is
  natural to get inspired by the methods developed over decades in
  classical functional test in close interaction with chip-design and
  manufacturing and start to see quantum device
  characterization from the perspective of this decision problem. This
  will become even more urgent with increasing numbers of qubits and
  mandatory if in the future mass production of quantum devices is to
  become a reality. \\
  
  First 
  steps in this direction of change of perspective of quantum device characterization were taken in \cite{Mil.Bra.Gir:19}, where runtime
  statistics of measurement sequences were introduced allowing one to
  certify entanglement of states and channels with limited data, based
  on the method of truncated moment sequences
  \cite{PhysRevA.96.032312,PhysRevA.102.052406}, and in \cite{PhysRevA.108.022602} 
  with the introduction of appropriate decision criteria in Bayesian metrology. However, these
  techniques are limited to a few qubits.  Here we go a step
  further and adapt classical functional memory testing based on patterns \cite{gupta_semiconductor_2009} to
  functional testing of quantum memories. Pattern-based memory testing
  writes many patterns of 0s and 1s into the
  memory and then reads them out again. The results of entire chips or
  chip areas are represented with bit-failure maps that can give
  immediate information on failure mechanisms, such as failing bit-lines or
  word-lines, stuck memory cells, or inhomogeneous wafer processing.  A
  large plethora of different patterns was developed in parallel with
  processing technology and chip design in order to make the chip fail as
  quickly as possible. Inversely, if a chip passes a large number of
  such specialized patterns without error, the probability of
  overlooking failing cells becomes exceedingly small \cite{gupta_semiconductor_2009}.\\
  In adapting the pattern approach to quantum memories, it is clear that
  the set of patterns has to be substantially extended in order to
  include specificities of quantum systems such as coherent
  superpositions and entanglement.  Instead of just 0s and 1s, we can
  now write additional combinations of other single-qubit states into
  the quantum memory, such as eigenstates of the Pauli operators
  $\sigma_x$ and $\sigma_y$.  In addition, there are new failure
  mechanisms such as decoherence, couplings to parasitic two-level
  fluctuators, unintended couplings to other qubits, or frequency
  collisions for which new patterns need to be designed that discover
  and visualize the corresponding failures efficiently. We expect that
  quantum functional testing will become an own discipline with
  quantum patterns designed in parallel to the hardware development,
  just as in the classical case.  Here we go a first step in that
  direction by introducing a set  of classically inspired patterns,
  which already offer efficient ways to get new insight into the functioning of
  two of IBM's recent quantum processors. 

\section{Results and Discussion}
\subsection{Classical Patterns}\label{sec:Classical Patterns}
\begin{figure}
	\includegraphics[width=\linewidth]{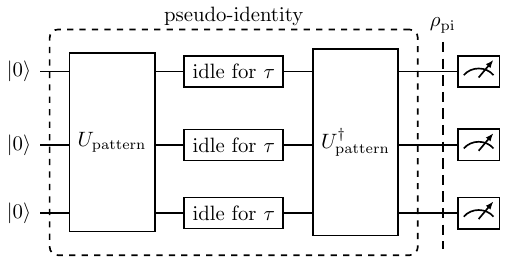}
	\caption{\textbf{Quantum circuit for pattern-based functional testing}. After qubits are initialized in state $\ket{0}$, a unitary operation $U_{\text{pattern}}$, depending on the intended pattern, is applied. After some time $\tau$, during which the qubits have been left idling, the inverse unitary $U_{\text{pattern}}^\dagger$ is applied.
  This implements a pseudo-identity operation designed to return the qubits to state $\ket{0}$ on an ideal, error-free quantum processor. However, on a real device, the state $\rho_{\text{pi}}$ after the pseudo-identity operation will differ from $\ket{0}$. Finally, all qubits are measured in the computational basis, allowing calculation of the fidelity $ F(\ketbra{0}{0}, \rho_{\text{pi}})$.
	}
	\label{fig:1}
\end{figure}

Blank patterns, among the simplest techniques in classical memory testing, involve uniformly writing in and reading out either 0s or 1s across all memory cells. These patterns serve as effective tools for identifying a variety of errors, including stuck-at faults where a cell remains fixed at 0 or 1, transition faults hindering state changes during write operations, and data retention faults causing a memory cell to lose its logic value over time.

Similar to retention faults, a qubit in its excited state $\ket{1}$ can transition to its ground state $\ket{0}$ over time due to energy relaxation. This transition can be characterized by the relaxation time $T_1$. To measure $T_1$, we bring the qubit to its excited state by applying an X-gate, wait for some time $\tau$, then apply a second X-gate and measure. This implements a pseudo-identity operation that should return the state to $\ket{0}$ on an ideal quantum processor without any errors. On a real device, the state $\rho_{\text{pi}}$ after the pseudo-identity operation will differ from $\ket{0}$, see Figure~\ref{fig:1}. This deviation can be quantified using the fidelity $F(\ketbra{0}{0},\rho_{\text{pi}} )$ (see Section~\ref{sec:Methods} for details). We repeat this for various time intervals $\tau$ to see the exponential decrease in fidelity over time. Doing this for all qubits in parallel gives what we call a blank $\ket{1}$ pattern. 

\begin{figure}
	\includegraphics[width=\linewidth]{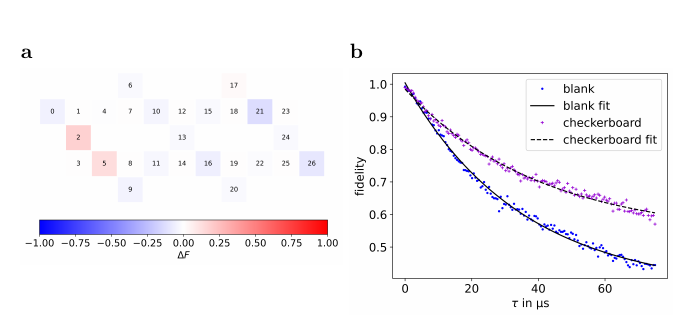}
	\caption{\textbf{Impact of states of adjacent qubits on relaxation time $T_1$}.
	Here we compare relaxation times of individual qubits on \textit{ibmq\_ehningen} under two conditions: blank and checkerboard patterns. In the blank setup, a $T_1$-experiment is performed on all qubits in parallel. In contrast, the checkerboard pattern divides qubits into alternating groups where two qubits in the same group are never adjacent, allowing to study how nearby qubit states affect relaxation times. Here we compare the checkerboard pattern A (for further details see Supplemental Material Section~\ref{sec:Hardware}) of $\ket{0}$ and $\ket{1}$ to the blank pattern of $\ket{1}$. By comparing fidelity $\Delta F = F_{\text{blank}}-F_{\text{checkerboard}}$  between the two setups ($\Delta F$ for spectator qubits in the checkerboard pattern is set to 0), differences hint at influences of neighboring qubits on relaxation. \textbf{a} Qubit failure map at $\tau = 75 \,\mathrm{\mu s}$. \textbf{b} Exponential decay of qubit 21's fidelity with relaxation times $T_{1,\text{checkerboard}}= 40 \,\mathrm{\mu s}$ and $T_{1,\text{blank}}= 35 \,\mathrm{\mu s}$.
	}
	\label{fig:2}
\end{figure}

Other patterns in classical memory include checkerboard patterns of 0s and 1s which are used to detect unwanted influences of neighboring cells.
To see if adjacent qubits affect relaxation, we use checkerboard patterns of $\ket{0}$s and $\ket{1}$s. We partition the qubits on a quantum chip into two distinct subgroups such that qubits within the same subgroup are never adjacent. To the set of qubits on which an experiment is performed we refer to as 'target qubits' while those that are not being investigated are referred to as 'spectator qubits'. To clarify which qubits are the target qubits in a given pattern we refer to checkerboard pattern A (or B) as defined in Supplemental Material Section~\ref{sec:Hardware}. Unless specified otherwise, all experiments were conducted using 2500 shots. Further information can be found in the Methods Section~\ref{sec:Methods}.
Comparing qubit behavior in blank and checkerboard patterns, shown in Figure~\ref{fig:2}, can reveal qubits, i.e. qubit 21 on \textit{ibmq\_ehningen}, where neighboring qubit states influence the relaxation time. It is apparent that energy relaxation is influenced by the state of adjacent qubits. $T_1$-times tend to be larger for most qubits in the checkerboard pattern, where neighboring qubits are left in the state $\ket{0}$. Qubits 2 and 5 of \textit{ibmq\_ehningen} in Figure~\ref{fig:2} are an exception to this rule. For those a possible mechanism is due to a frequency collision on the qubit group 2, 3, and 5, which will be discussed in Section~\ref{sec:Other hardware motivated patterns} below.

\begin{figure}
	\includegraphics[width=\linewidth]{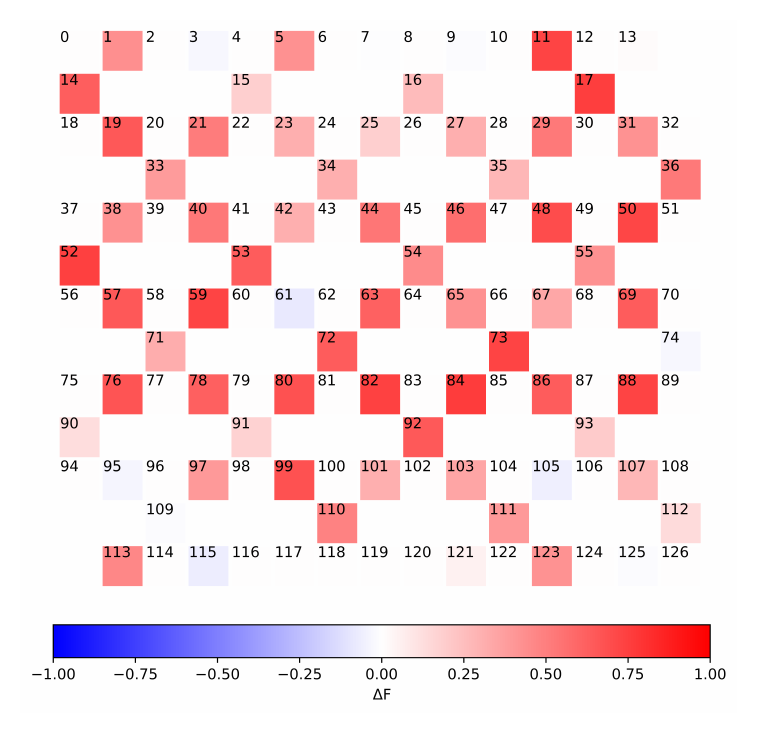}
	\caption{\textbf{Qubit failure map showing the influence of adjacent qubit dynamics on relaxation}.
	Comparison between a blank $\ket{1}$ pattern and a checkerboard $\ket{1}$ pattern A (see Supplemental Material Section~\ref{sec:Hardware}) with repeated application of an even number $N$ of X-gates to spectator qubits, where $\Delta F = F_{\text{blank}}-F_{\text{checkerboard with }X^N}$ on \textit{ibmq\_brisbane}. $\Delta F$ for spectator qubits in the checkerboard pattern A is set to 0 for clarity. Shown is the qubit failure map at $\tau = 50\,\mathrm{\mu s}$ with $N=250$ applied X-gates.
	}
	\label{fig:3}
\end{figure}

Taking it one step further, we introduce dynamics to the spectator qubits. Instead of leaving them idle in state $\ket{0}$, we repeatedly apply an even number of X-gates to induce excitation and de-excitation. A circuit for such a pattern can be seen in Supplemental Material Section~\ref{sec:Circuits}. The comparison between this modified pattern and a checkerboard pattern with idle spectator qubits is illustrated in Figure~\ref{fig:3}. Here, one can observe that the dynamics in neighboring qubits lead to faster relaxation.
One possible mechanism for the influence of switching neighboring qubits is heating, caused by excitation and de-excitation of qubits, leading to an increased quasiparticle density which has previously been linked to decreased $T_1$-times \cite{Wang2014,Pan2022}.

\subsection{Superposition Patterns}\label{sec:Superposition Patterns}
Unlike in classical memory testing, qubits can also be prepared in superposition states such as an eigenstate of $\sigma_x$, i.e. $\ket{+} =(\ket{0}+\ket{1})/\sqrt{2}$. The timescale on which qubits in a superposition state lose their well-defined phase and transition to a mixed state is given by $T_2$.
A qubit in the state $\ket{0}$ can be brought into the superposition state $\ket{+}$ by applying a Hadamard gate. After allowing it to idle for a duration $\tau$, a second Hadamard gate is applied, followed by measurement. In such an experiment, the probability of measuring the qubit in $\ket{0}$ decays exponentially with time $\tau$. Performing this procedure parallelly on all qubits on a chip creates what we refer to as a blank $\ket{+}$ pattern.
By inserting an X-gate at $t=\tau/2$, known as a Spin- or Hahn-echo \cite{Hahn1950}, sensitivity to quasistatic, low-frequency contributions to dephasing can be reduced, and rotations due to frequency detuning can be refocused. Performing this procedure in parallel on all qubits on a chip results in what we call an {\em echoed} blank $\ket{+}$ pattern. The corresponding circuits are detailed in Supplemental Material Section~\ref{sec:Circuits}.

\begin{figure}
	\includegraphics[width=\linewidth]{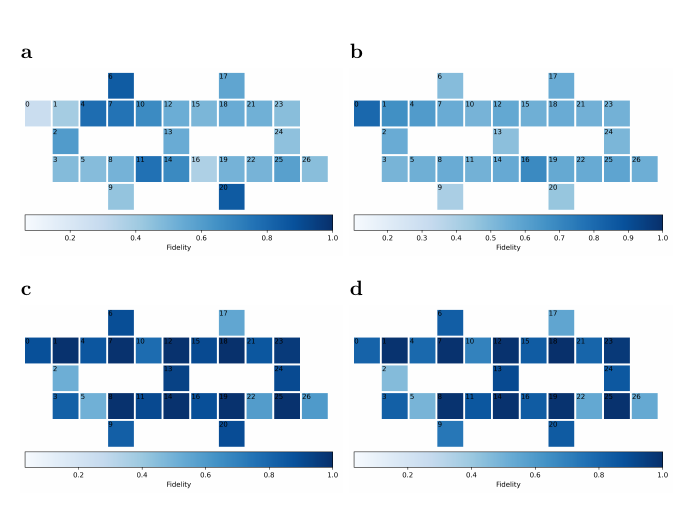}
	\caption{\textbf{Qubit failure maps showing the influence of adjacent qubit states on coherence}.  
	The Figure depicts the results of two distinct patterns: the echoed blank $\ket{+}$ pattern (\textbf{a} $\tau = 40 \,\mathrm{\mu s}$ and \textbf{b} $\tau = 75 \,\mathrm{\mu s}$) and the echoed checkerboard $\ket{+}$ pattern A with spectator qubits in $\ket{0}$ (\textbf{c} $\tau = 40 \,\mathrm{\mu s}$ and \textbf{d} $\tau = 75 \,\mathrm{\mu s}$). Qubits in the checkerboard pattern demonstrate notably longer coherence times compared to those in the blank pattern. 
	Some qubits in the blank pattern show a recovery of fidelity with time. Specifically, when examining qubit 0 in the blank $\ket{+}$ pattern, the fidelity is initially low at $\tau = 40 \,\mathrm{\mu s}$, but it notably improves at $\tau = 75 \,\mathrm{\mu s}$. This phenomenon is not observed for the checkerboard $\ket{+}$ pattern.
	}
	\label{fig:4}
\end{figure}

In Figure~\ref{fig:4} \textbf{a} and \textbf{b} we present the qubit failure map of an echoed blank $\ket{+}$ pattern at $\tau = 40 \,\mathrm{\mu s}$ and $\tau = 75 \,\mathrm{\mu s}$, while in Figure~\ref{fig:4} \textbf{c} and \textbf{d} we display the failure map for the echoed checkerboard $\ket{+}$ pattern with spectator qubits in $\ket{0}$. For both cases, one expects a decrease of fidelity over time, occurring on the same timescale. However, qubits in the checkerboard pattern exhibit notably longer coherence times compared to the blank pattern. Interestingly, there is a fidelity recovery at longer times for some qubits in the blank pattern.
For example, when observing qubit 0 in the echoed blank $\ket{+}$ pattern, the fidelity is notably low at $\tau = 40 \,\mathrm{\mu s}$, but it significantly improves at $\tau = 75 \,\mathrm{\mu s}$. This fidelity recovery of some qubits is not observed in an echoed checkerboard $\ket{+}$ pattern.
By comparing different patterns it is possible to detect unwanted influences of neighboring qubits on coherence, negatively impacting the ability of qubits to store quantum states as intended.

\begin{figure}
	\includegraphics[width=\linewidth]{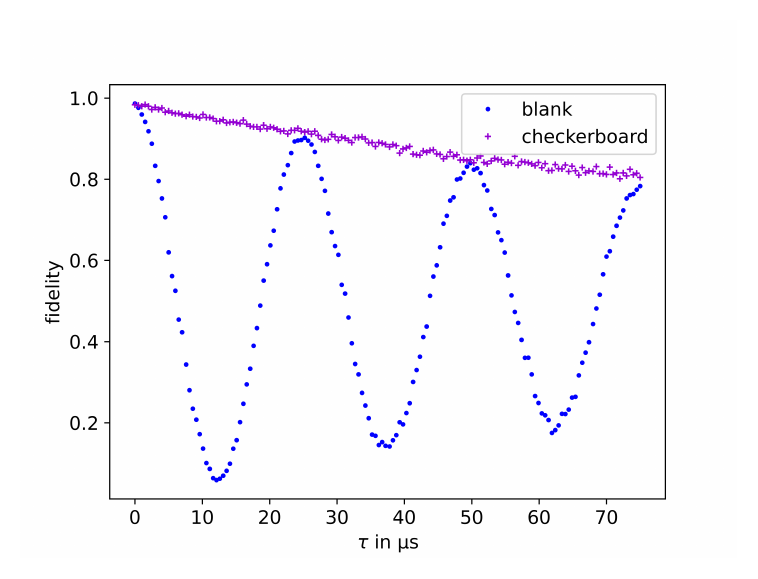}
	\caption{\textbf{Influence of adjacent qubit states on fidelity of superposition states}.  
	Comparing data of qubit 0 from two $\ket{+}$ patterns performed on \textit{ibmq\_ehningen}. In the case of the echoed checkerboard $\ket{+}$ pattern, the probability to measure 0 decays exponentially, as expected. On the other hand, if the experiment is performed on all qubits simultaneously as in the echoed blank $\ket{+}$ pattern one can see oscillations in the probability.}
	\label{fig:5}
\end{figure}

From the example of qubit 0 on \textit{ibmq\_ehningen} in Figure~\ref{fig:5} we can see that in the echoed checkerboard $\ket{+}$ pattern, fidelity decays exponentially as expected. However, data obtained from an echoed blank $\ket{+}$ pattern for qubit 0 displays an oscillatory behavior.

\begin{figure}
	\includegraphics[width=\linewidth]{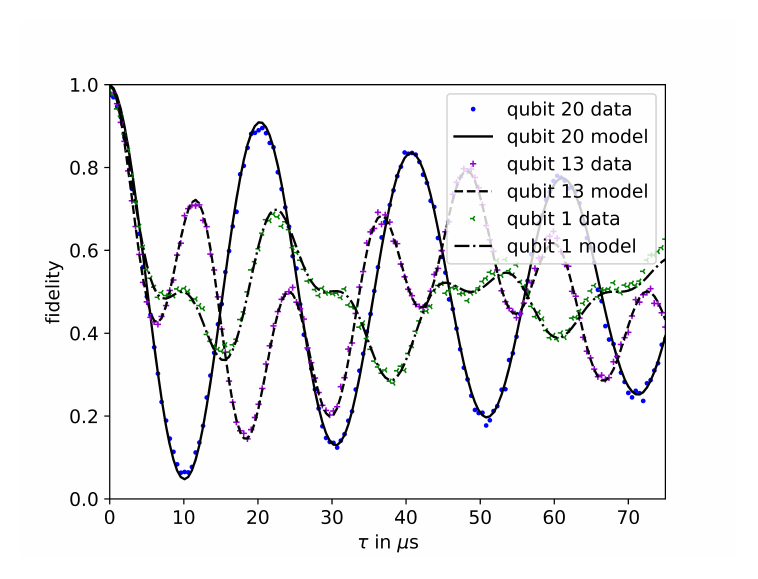}
	\caption{\textbf{Fidelity of qubits with different numbers of nearest neighbors from an echoed blank $\ket{+}$ pattern.}  
  Analysis of fidelity oscillations over delay time was performed for three specific qubits on \textit{ibmq\_ehningen}. The complexity of these oscillations varies with the number of adjacent qubits. Qubit 20, with only one nearest neighbor (qubit 19), exhibits simpler oscillations compared to qubit 13, with neighbors 12 and 14, and qubit 1, with neighbors 0, 2, and 4.
  The data was modeled using the Hamiltonian described in Equation (\ref{eq:zz_Hamiltonian}). As an example, for qubit 20, the Hamiltonian is given by $H /\hbar = \Omega_{zz}^1\sigma_z^{q20}\sigma_z^{q19}$. The fitting process employed the Lindblad Master Equation, as detailed in  Section~\ref{sec:Simulations of patterns} of the Supplemental Material. The resulting fitted parameters are listed in Table~\ref{tab:1}.
  } 
	\label{fig:6}
\end{figure}

Figure~\ref{fig:6} displays the fidelity as a function of $\tau$ from an echoed blank $\ket{+}$ pattern for three qubits of \textit{ibmq\_ehningen}: qubit 20 with one nearest neighbor, qubit 13 with two nearest neighbors, and qubit 1 with three nearest neighbors. As the number of adjacent qubits increases, the oscillation transitions from sinusoidal to an oscillation with more than a single frequency.

Multi-frequency oscillations in qubit observables are indicative of non-Markovian dynamics in both idle and driven qubits \cite{Burkard2023,Agarwal2023}. These non-Markovian dynamics can be caused by interactions of qubits with two level systems in their environment \cite{Agarwal2023}. Such two level systems can emerge from defects in the device materials and are also known as a leading cause of decoherence and energy relaxation of superconducting qubits \cite{Lisenfeld2019,Mueller2015,Mueller2019}.
While $\langle \sigma_x \rangle$ of qubits 12 and 1 in Figure~\ref{fig:6} exhibits multi-frequency oscillations, the observed behavior can also be explained by interactions between physically adjacent qubits. In this scenario, neighboring qubits function as an environment with memory, leading to non-Markovian dynamics when examining individual qubits. Such crosstalk effects have been attributed to a residual $zz$-coupling of neighboring qubits \cite{Perrin2024,Gambetta2012,Heunisch2023,Sung2021,Murali2020}. Taking all coupled qubits into account, the evolution is no longer non-Markovian.

To explain the oscillations shown in Figure~\ref{fig:6}, we adopt the model discussed in \cite{Samach2022} and \cite{OMalley2015}. As illustrated by the simulations in Supplemental Material Section~\ref{sec:Simulations of patterns}, a coupling of the form $H/\hbar = \Omega_{zz} \sigma_z \otimes \sigma_z $ between adjacent qubits can explain the observed phenomena. This model can be straightforwardly extended to include interactions with $N$ nearest neighbors.

\begin{align}\label{eq:zz_Hamiltonian}
	\frac{H}{\hbar} = \sum_{i=1}^{N} \Omega_{zz}^i \sigma_z^1  \sigma_z^{i+1}
\end{align}

Figure~\ref{fig:6} shows that this model still explains the data for two or three coupled qubits very well. Estimated coupling strengths $\Omega_{zz}$ obtained via a least squares fit are displayed in Table~\ref{tab:1}. Oscillations resulting from such couplings do not manifest in the patterns of states $\ket{0}$ and $\ket{1}$ (see Figure~\ref{fig:2}), as these states are eigenstates of $\sigma_z$.

\begin{table}[ht]
	\begin{center}
		\begin{tabular}{lcccc}
			Qubit & $\Omega_{zz}^1$ & $\Omega_{zz}^2$ &  $\Omega_{zz}^3$ & $T_2$\\
			\hline
			20 & 0.155 & - & - & 204 \\
			13 & 0.163 & 0.097 & -& 180 \\
			1 & 0.126 & 0.081 & 0.081 & 94\\
			\hline
		\end{tabular}
	\end{center}
	\caption{Examples for estimated $zz$-coupling strengths $\Omega_{zz}^{i}$ in $2\pi \cdot \mathrm{MHz}$ and decoherence times $T_2$ in $\mathrm{\mu s}$ of qubits on \textit{ibmq\_ehningen}.}
	\label{tab:1}
	\end{table}

\subsection{Entangled patterns}\label{sec:Entangled patterns}

\begin{figure}
	\includegraphics[width=\linewidth]{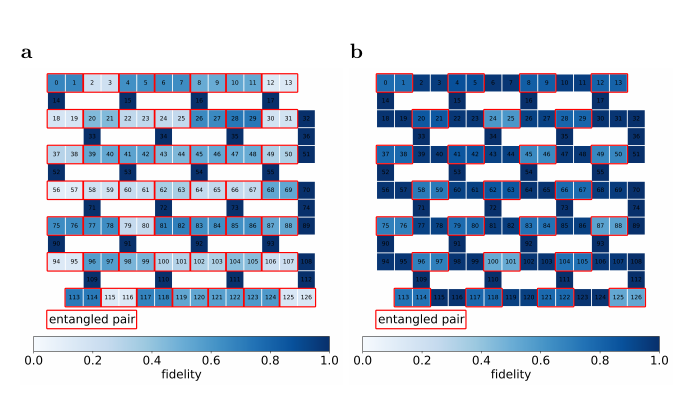}
	\caption{\textbf{Fidelities of entangled states in $\ket{\Phi^-}$ patterns}.
	We compare data from two echoed $\ket{\Phi^-}$ patterns at $\tau = 21 \,\mathrm{\mu s}$ on \textit{ibm\_brisbane}. \textbf{a} Pattern with most qubits in a $\ket{\Phi^-}$ state  \textbf{b} Pattern with pairs of qubits in $\ket{\Phi^-}$ where qubit pairs are always separated by at least one qubit in $\ket{0}$. Red boxes indicate entangled qubit pairs. Qubits not inside a red box are spectators, left idle in $\ket{0}$, and their fidelity is calculated individually. Qubits in the pattern where qubits are separated by one spectator in $\ket{0}$ show notably longer coherence times compared to those in the blank $\ket{\Phi^-}$ pattern. $\ket{\Phi^-}$-states are affected by the residual $zz$-coupling between adjacent qubits as also discussed in case of $\ket{+}$-states.}
	\label{fig:7}
\end{figure}

Entanglement plays a crucial role in quantum algorithms \cite{Jozsa03}. Hence, we investigate how long an entangled state can be maintained within a quantum device. For instance, let's examine one of the two-qubit Bell-states $\ket{\Phi^\pm}=\left(\ket{0}\otimes\ket{0}\pm\ket{1}\otimes\ket{1}\right)/\sqrt{2}$. To investigate such a state, we employ a circuit illustrated in Supplemental Material Section~\ref{sec:Circuits}, with varying duration $\tau$.
Figure~\ref{fig:7} shows a comparison between two patterns involving entangled states: one where between every pair of entangled qubits, at least one qubit remains in state $\ket{0}$, and another where most qubits are entangled pairwise in the $\ket{\Phi^-}$-state, so that neighboring pairs are next to each other. In both cases, a decrease in fidelity over time is expected at the same timescale. However, qubits in the second pattern exhibit notably longer coherence times compared to those in the blank $\ket{\Phi^-}$ pattern. Similar to superposition states, entangled states are impacted by the previously discussed residual $zz$-coupling between adjacent qubits, significantly reducing the lifespan of entangled states if adjacent qubits do not remain in the state $\ket{0}$.

\begin{figure}
	\includegraphics[width=\linewidth]{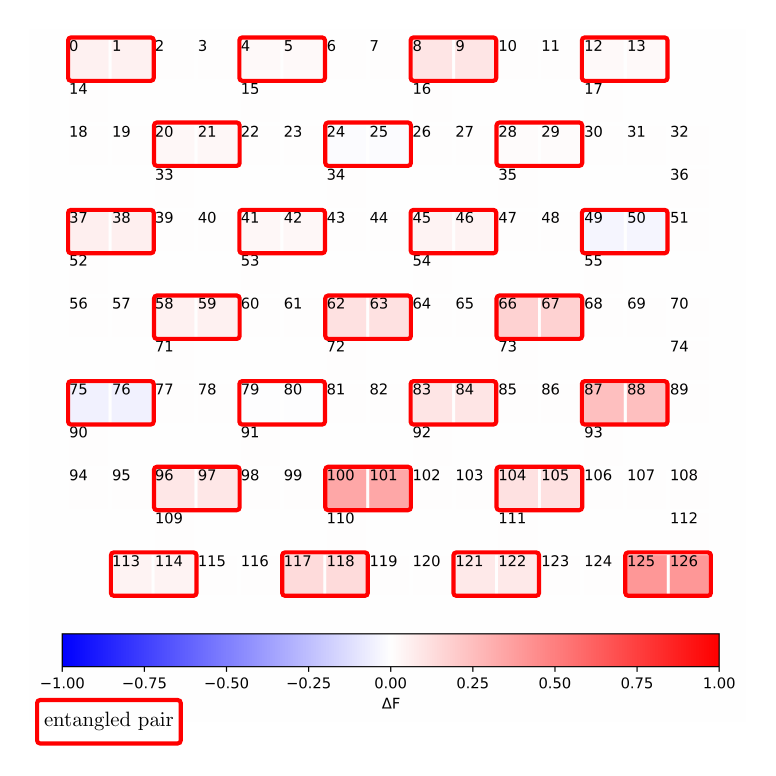}
	\caption{\textbf{Qubit failure map comparing fidelities between a pattern with $\ket{\Phi^+}$ entangled states and checkerboard $\ket{+}$ patterns}.
	Here fidelity of an echoed $\ket{\Phi^+}$ pattern is compared with the fidelity of two echoed checkerboard $\ket{+}$ patterns on \textit{ibm\_brisbane}.
  For the comparison, we first calculated the fidelity of the qubit pairs (red boxes) in the $\ket{\Phi^+}$ pattern. Then, we determined the fidelity of all target qubits in checkerboard pattern A and checkerboard pattern B. Finally, we obtained the joint fidelity of the qubit pairs, entangled in case of the $\ket{\Phi^+}$ pattern, by multiplying the fidelities from the two checkerboard patterns A and B of the individual qubits in the same pairs. Shown is $\Delta F = F_{\ket{+}} - F_{\ket{\Phi^+}}$ at $\tau = 50\,\mathrm{\mu s}$. The fidelity difference $\Delta F$ of all qubits that have not been entangled in case of the $\ket{\Phi^+}$ pattern is set to 0.}
	\label{fig:8}
\end{figure}

In Figure~\ref{fig:8}, we compare the fidelities of a pattern with entangled states $\ket{\Phi^+}$ to those of checkerboard $\ket{+}$ patterns. In this case we found that the entangled pattern is more sensitive to noise, a result that also holds true for comparisons with other Bell 2-qubit states.

\subsection{Other hardware motivated patterns}\label{sec:Other hardware motivated patterns}

Frequency collisions among neighboring qubits in fixed-frequency transmon systems can lead to unintentional driving of individual qubits or interactions between qubits and are a potential cause of crosstalk effects \cite{Ketterer2023}. For instance, if the frequencies of adjacent qubits $A$ and $B$ are close, i.e. $\omega_{01}^A \approx \omega_{01}^B $, operations on qubit $A$ can unintentionally affect qubit $B$. While avoiding these straightforward collisions is simple, there are more intricate resonances, like those involving transitions to higher-energy levels, beyond the main computational space \cite{Hertzberg2021}.

\begin{figure}
	\includegraphics[width=\linewidth]{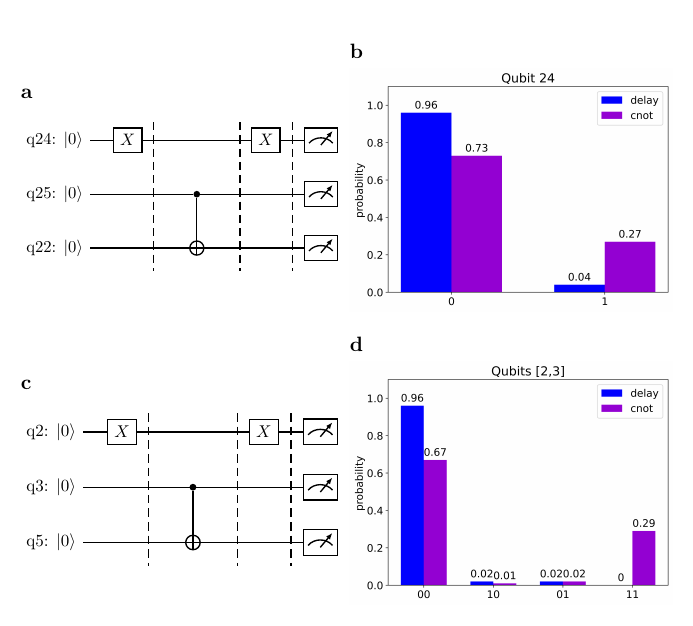}
	\caption{\textbf{Failure of qubit 24 due to frequency collision}.
  The circuits in \textbf{a} and \textbf{c} are compared to a circuit where, instead of applying a CNOT gate, all three qubits are left idle for a duration of $\tau = 500 \, \mathrm{ns}$. As shown in the histogram in \textbf{b}, the probability of measuring qubit 24 in the state $\ket{0}$ is significantly lower when a CNOT gate is applied to its neighbors compared to when the CNOT is replaced with a delay of the same duration. This is due to a frequency collision of the form $\omega_{12}^{q24} = 4.7329 \,\mathrm{GHz} \approx 4.7251 \,\mathrm{GHz} = \omega_{01}^{q22}$. Similarly, the histogram in \textbf{d} shows that the probability of measuring qubits 2 and 3 in the state $\ket{0}$ is also significantly lower when a CNOT gate is applied. This can be explained by a frequency collision of the form $\omega_{02}^{q3} - \omega_{01}^{q2} = 5.0709 \,\mathrm{GHz} \approx 5.0712 \,\mathrm{GHz} = \omega_{01}^{q5}$.
  The circuits were executed on \textit{ibmq\_ehningen} with 10000 shots, so shot noise is insignificant. 
	}
	\label{fig:9}
\end{figure}

On \textit{ibmq\_ehningen} a CNOT gate is realized using a cross-resonance gate, where the control qubit is driven with the transition frequency of the target qubit \cite{Krantz2019}. In some cases this can lead to unwanted resonances during the application of CNOT gates.
Here we investigate qubit failures due to frequency collisions involving three qubits $A$, $B$ and $C$. 
Qubit $A$ is first brought to its state $\ket{1}$ through the application of an X-gate. Subsequently, a CNOT gate is employed on a neighboring qubit pair $B$ and $C$, where $B$ is adjacent to $A$ and acts as the control qubit, with $C$ as the target qubit. Following this, a second X-gate is applied to qubit $A$ prior to the measurement of all qubits involved.
This process is depicted in Figure~\ref{fig:9} \textbf{a} for the qubit triplet 24, 25 and 22. For these three qubits we have that $\omega_{12}^{q24} \approx \omega_{01}^{q22}$.The qubit couplings form a chain: 24 - 25 - 22. Hence, when we drive qubit 25 with the transition frequency of qubit 22 to implement a CNOT gate with qubit 25 as the control qubit and qubit 22 as the target qubit, this can cause a transition of qubit 24 to its second excited state $\ket{2}$, as is well illustrated in Figure 2 of \cite{Ketterer2023}. The following X-gate then cannot bring qubit 24 back to state $\ket{0}$. Additionally, in the standard measurement procedure used in IBM devices, a qubit in state $\ket{2}$ will be measured as if it were in state $\ket{1}$ resulting in a measurement with an increased probability to measure qubit 24 in $\ket{1}$. We compare the measurement results from the circuit shown in Figure~\ref{fig:9} \textbf{a} with those from a circuit where the CNOT gate is replaced by a delay of equal duration. The histogram in Figure~\ref{fig:9} \textbf{b} illustrates the decrease in fidelity of qubit 24 when a CNOT gate is applied to its neighbors.

\begin{figure}
	\includegraphics[width=\linewidth]{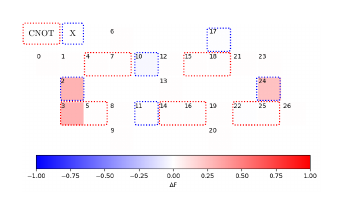}
	\caption{\textbf{Failure map of \textit{ibmq\_ehningen} due to frequency collisions}.
	The chip is partitioned into sets of qubit triplets and fidelity is compared between a circuit where a CNOT gate is applied to two qubits after an X-gate was applied to a neighboring qubit and one where after the application of the first X-gate the qubits have been left idle for the duration a CNOT gate would have taken: $\Delta F = F_{\text{delay}} - F_{\text{CNOT}}$.
	}
	\label{fig:10}
\end{figure}

By partitioning the quantum chip into groups of three qubits, one can create a pattern that allows simultaneous testing of multiple qubit triplets. Each triplet is separated from the others by at least one qubit, which remains idle in $\ket{0}$. Figure~\ref{fig:10} illustrates a pattern comparing the fidelity of circuits with and without CNOT gates and revealing two triplets with frequency collisions one of which is the discussed case of the qubit triplet 24, 25 and 22.

The other directly visible case of frequency collision involves the qubit triplet 2, 3, 5, where the fidelity of qubits 2 and 3 decreases due to a frequency collision of type $\omega_{02}^{q3} - \omega_{01}^{q2} \approx  \omega_{01}^{q5} $. If qubit 2 is in the state $\ket{1}$ and qubit 3 in $\ket{0}$ in a setup like in Figure~\ref{fig:9} \textbf{c}, the cross-resonance pulse at frequency $\omega_{01}^{q5}$ applied to qubit 3, which is intended to generate a CNOT gate between qubits 3 and 5, also renders the combined transition q3: $0 \rightarrow 2 $ and q2: $1 \rightarrow 0 $ resonant as displayed in Figure~\ref{fig:9} \textbf{d}. As a result, it also implements an entangling gate between qubits 2 and 3, resulting in a state of the form $\ket{\Psi} = \alpha \ket{0}_{q2}\ket{2}_{q3} + \beta \ket{1}_{q2}\ket{0}_{q3}$. 
Applying an X-gate to qubit 2 before measurement results in a state $\ket{\Psi} = \alpha \ket{1}_{q2}\ket{2}_{q3} + \beta \ket{0}_{q2}\ket{0}_{q3}$. Therefore, both qubits are measured either in state $\ket{1}$ with probability $|\alpha|^2$ or in state $\ket{0}$ with probability $|\beta|^2$ as is displayed in Figure~\ref{fig:9}.

The frequency collision involving qubits 2, 3, and 5 may also provide an explanation for the different sign of $\Delta F=F_{\text{blank}}-F_{\text{checkerboard}}$ observed for qubits 2 and 5 in Figure~\ref{fig:2}. Since $\omega_{02}^{q3} \approx \omega_{01}^{q2} + \omega_{01}^{q5} $, the frequency collision only occurs if qubit 3 is prepared in state $\ket{0}$ (checkerboard pattern).

\section{Methods}\label{sec:Methods}

Instead of testing every possible state of a quantum information processing device, the focus is on identifying known error mechanisms by designing patterns that test for these errors simultaneously. Our pattern-based quantum memory testing procedure, illustrated in Figure~\ref{fig:1}, starts by encoding a specific state into the memory using unitary gates applied to the qubits according to specific patterns.
Then, we allow the qubits to idle for a duration of time $\tau$ before applying the inverse operations and performing measurements. In an ideal, noise-free scenario, these operations would result in the identity operation. However, in the presence of noise, the operations may induce unintended dynamics. By comparing the expected measurement results with the actual probability distribution, we aim to gain insights into error mechanisms. 

For patterns involving $\ket{+}$- or $\ket{-}$-states, sensitivity to low-frequency noise can be mitigated by placing an X-gate at $\tau/2$ for the target qubits. This technique is known as Spin- or Hahn-echo \cite{Hahn1950} and, consequently, we refer to such patterns as echoed ones.
One common error mechanism involves unwanted interactions with adjacent qubits, where changing the state of one qubit might unintentionally affect the state of a nearby qubit. Alongside patterns with Bell states, checkerboard patterns of $\ket{0}$s and $\ket{1}$s or  $\ket{+}$ states and $\ket{0}$s are employed to identify these errors.
Patterns containing entangled states with many qubits are potentially very useful to detect specific error scenarios. However, achieving high-fidelity entanglement with more than a handful of qubits remains difficult in current devices, leading to predominantly noisy measurements as can be seen from Supplemental Material Section~\ref{sec:Size limit of entangled states}. Three qubit patterns, introduced in Section~\ref{sec:Other hardware motivated patterns} are used to find qubit failures due to frequency collisions. 

We employed our method on IBM's superconducting fixed-frequency transmon qubits, utilizing two quantum computers: the 27-qubit device \textit{ibmq\_ehningen} (Falcon r5.11 architecture) and the 127-qubit device \textit{ibm\_brisbane} (Eagle r3 architecture), with layouts as illustrated in Section~\ref{sec:Hardware} of the Supplemental Material. 
Measurements on \textit{ibmq\_ehningen} were conducted in January 2024, and measurements on \textit{ibm\_brisbane} took place between October 2023 and May 2024. Unless otherwise specified, all experiments were performed with 2500 shots. Quantum circuits were constructed using Qiskit 0.45.1, and gate executions were calibrated according to IBM's specifications.

As a metric for measuring the closeness of two quantum states $\rho_0$ and $\rho_{\text{pi}}$ we use the fidelity $F(\rho_0,\rho_{\text{pi}}) = \tr(\rho_0\rho_{\text{pi}})$. In our scenario, the initial state of a qubit is always the state  $\rho_0= \ketbra{0}{0}$ and thus the fidelity is the probability to measure the qubit in $\ket{0}$.
For $N$-qubit states the fidelity $F(\ket{0}^{\otimes N}\bra{0}^{\otimes N}, \rho_{\text{pi}})$ is the joint probability to measure all $N$ qubits in state $\ket{0}$.
We use $N=1$ for single qubit states in blank and checkerboard patterns, for the patterns to test for frequency collisions in Section~\ref{sec:Other hardware motivated patterns} and always for calculating the fidelity of spectator qubits. 
For patterns with two-qubit entangled states, we use $N=2$.

\section{Summary}
We proposed a pattern-based approach to quantum functional testing, with the objective to determine as quickly as possible and with a high level of confidence whether a device is malfunctioning or operating correctly. Using patterns crafted specifically to stress a device based on known error mechanisms, we showed that our method is effective in inducing rapid failures. For instance, we developed three-qubit patterns to detect qubit failures caused by frequency collisions, or single-qubit superposition patterns to expose unintended couplings between qubits. Qubit-failure maps allow us to visualize failure mechanisms of entire chips and promptly locate errors. Expanding upon these tailored patterns, one can also incorporate numerous random patterns, including various entangled states of different complexities. This enables the discovery of previously unknown failure mechanisms. The probability of overlooking hardware failures is expected to decrease exponentially with the number of independent patterns, offering a practical and scalable solution for certifying device functionality.

\section*{Acknowledgements}
This work is funded by the Ministry of Economic Affairs, Labor and Tourism Baden-W\"urttemberg in the frame of the Competence Center Quantum Computing Baden-W\"urttemberg (project ``QORA II'').

We acknowledge funding from the Deutsche Forschungsgemeinschaft (DFG, German Research Foundation) through the Research Unit FOR 5522/1, Grant No. 499180199.

\section*{Data Availability}
The data that support the findings of this study are available from the corresponding author upon reasonable request.

\bibliographystyle{naturemag}
\bibliography{mybibs_bt}

\begin{thebibliography}{10}
\expandafter\ifx\csname url\endcsname\relax
  \def\url#1{\texttt{#1}}\fi
\expandafter\ifx\csname urlprefix\endcsname\relax\def\urlprefix{URL }\fi
\providecommand{\bibinfo}[2]{#2}
\providecommand{\eprint}[2][]{\url{#2}}

\bibitem{chuang_prescription_1997}
\bibinfo{author}{Chuang, I.~L.} \& \bibinfo{author}{Nielsen, M.~A.}
\newblock \bibinfo{title}{Prescription for experimental determination of the dynamics of a quantum black box}.
\newblock \emph{\bibinfo{journal}{Journal of Modern Optics}} \textbf{\bibinfo{volume}{44}}, \bibinfo{pages}{2455--2467} (\bibinfo{year}{1997}).
\newblock \urlprefix\url{http://www.tandfonline.com/doi/abs/10.1080/09500349708231894}.

\bibitem{PhysRevA.63.020101}
\bibinfo{author}{Fiur\'a\ifmmode~\check{s}\else \v{s}\fi{}ek, J.} \& \bibinfo{author}{Hradil, Z. c.~v.}
\newblock \bibinfo{title}{Maximum-likelihood estimation of quantum processes}.
\newblock \emph{\bibinfo{journal}{Phys. Rev. A}} \textbf{\bibinfo{volume}{63}}, \bibinfo{pages}{020101(R)} (\bibinfo{year}{2001}).
\newblock \urlprefix\url{https://link.aps.org/doi/10.1103/PhysRevA.63.020101}.

\bibitem{PhysRevLett.86.4195}
\bibinfo{author}{D'Ariano, G.~M.} \& \bibinfo{author}{Lo~Presti, P.}
\newblock \bibinfo{title}{Quantum tomography for measuring experimentally the matrix elements of an arbitrary quantum operation}.
\newblock \emph{\bibinfo{journal}{Phys. Rev. Lett.}} \textbf{\bibinfo{volume}{86}}, \bibinfo{pages}{4195--4198} (\bibinfo{year}{2001}).
\newblock \urlprefix\url{https://link.aps.org/doi/10.1103/PhysRevLett.86.4195}.

\bibitem{PhysRevLett.90.193601}
\bibinfo{author}{Altepeter, J.~B.} \emph{et~al.}
\newblock \bibinfo{title}{Ancilla-assisted quantum process tomography}.
\newblock \emph{\bibinfo{journal}{Phys. Rev. Lett.}} \textbf{\bibinfo{volume}{90}}, \bibinfo{pages}{193601} (\bibinfo{year}{2003}).
\newblock \urlprefix\url{https://link.aps.org/doi/10.1103/PhysRevLett.90.193601}.

\bibitem{PhysRevA.64.012314}
\bibinfo{author}{Childs, A.~M.}, \bibinfo{author}{Chuang, I.~L.} \& \bibinfo{author}{Leung, D.~W.}
\newblock \bibinfo{title}{Realization of quantum process tomography in nmr}.
\newblock \emph{\bibinfo{journal}{Phys. Rev. A}} \textbf{\bibinfo{volume}{64}}, \bibinfo{pages}{012314} (\bibinfo{year}{2001}).
\newblock \urlprefix\url{https://link.aps.org/doi/10.1103/PhysRevA.64.012314}.

\bibitem{leonhardt_tomographic_1995}
\bibinfo{author}{Leonhardt, U.}, \bibinfo{author}{Paul, H.} \& \bibinfo{author}{D’Ariano, G.~M.}
\newblock \bibinfo{title}{Tomographic reconstruction of the density matrix via pattern functions}.
\newblock \emph{\bibinfo{journal}{Physical Review A}} \textbf{\bibinfo{volume}{52}}, \bibinfo{pages}{4899--4907} (\bibinfo{year}{1995}).
\newblock \urlprefix\url{http://link.aps.org/doi/10.1103/PhysRevA.52.4899}.

\bibitem{kosut_quantum_2008}
\bibinfo{author}{Kosut, R.~L.}
\newblock \bibinfo{title}{Quantum {Process} {Tomography} via {L1}-norm {Minimization}}.
\newblock \emph{\bibinfo{journal}{0812.4323}}  (\bibinfo{year}{2008}).
\newblock \urlprefix\url{http://arxiv.org/abs/0812.4323}.

\bibitem{christandl_reliable_2012}
\bibinfo{author}{Christandl, M.} \& \bibinfo{author}{Renner, R.}
\newblock \bibinfo{title}{Reliable {Quantum} {State} {Tomography}}.
\newblock \emph{\bibinfo{journal}{Physical Review Letters}} \textbf{\bibinfo{volume}{109}}, \bibinfo{pages}{120403} (\bibinfo{year}{2012}).
\newblock \urlprefix\url{http://link.aps.org/doi/10.1103/PhysRevLett.109.120403}.

\bibitem{blume-kohout_robust_2012}
\bibinfo{author}{Blume-Kohout, R.}
\newblock \bibinfo{title}{Robust error bars for quantum tomography}.
\newblock \emph{\bibinfo{journal}{arXiv:1202.5270}}  (\bibinfo{year}{2012}).
\newblock \urlprefix\url{http://arxiv.org/abs/1202.5270}.
\newblock \bibinfo{note}{ArXiv: 1202.5270}.

\bibitem{rey-de-castro_time-resolved_2013}
\bibinfo{author}{Rey-de Castro, R.}, \bibinfo{author}{Cabrera, R.}, \bibinfo{author}{Bondar, D.~I.} \& \bibinfo{author}{Rabitz, H.}
\newblock \bibinfo{title}{Time-resolved quantum process tomography using {Hamiltonian}-encoding and observable-decoding}.
\newblock \emph{\bibinfo{journal}{New Journal of Physics}} \textbf{\bibinfo{volume}{15}}, \bibinfo{pages}{025032} (\bibinfo{year}{2013}).
\newblock \urlprefix\url{http://iopscience.iop.org/1367-2630/15/2/025032}.

\bibitem{ohliger_efficient_2013}
\bibinfo{author}{Ohliger, M.}, \bibinfo{author}{Nesme, V.} \& \bibinfo{author}{Eisert, J.}
\newblock \bibinfo{title}{Efficient and feasible state tomography of quantum many-body systems}.
\newblock \emph{\bibinfo{journal}{New Journal of Physics}} \textbf{\bibinfo{volume}{15}}, \bibinfo{pages}{015024} (\bibinfo{year}{2013}).
\newblock \urlprefix\url{http://iopscience.iop.org/1367-2630/15/1/015024}.

\bibitem{Emerson05.2}
\bibinfo{author}{Emerson, J.}, \bibinfo{author}{Alicksi, R.} \& \bibinfo{author}{\.{Z}yczkowski, K.}
\newblock \bibinfo{title}{Scalable noise estimation with random unitary operators}.
\newblock \emph{\bibinfo{journal}{J. Opt. B: Quantum semiclass. Opt.}} \textbf{\bibinfo{volume}{7}}, \bibinfo{pages}{S347--S352} (\bibinfo{year}{2005}).

\bibitem{knill_randomized_2008}
\bibinfo{author}{Knill, E.} \emph{et~al.}
\newblock \bibinfo{title}{Randomized benchmarking of quantum gates}.
\newblock \emph{\bibinfo{journal}{Physical Review A}} \textbf{\bibinfo{volume}{77}}, \bibinfo{pages}{012307} (\bibinfo{year}{2008}).
\newblock \urlprefix\url{https://link.aps.org/doi/10.1103/PhysRevA.77.012307}.

\bibitem{bendersky_selective_2009}
\bibinfo{author}{Bendersky, A.}, \bibinfo{author}{Pastawski, F.} \& \bibinfo{author}{Paz, J.~P.}
\newblock \bibinfo{title}{Selective and efficient quantum process tomography}.
\newblock \emph{\bibinfo{journal}{Physical Review A}} \textbf{\bibinfo{volume}{80}}, \bibinfo{pages}{032116} (\bibinfo{year}{2009}).
\newblock \urlprefix\url{https://link.aps.org/doi/10.1103/PhysRevA.80.032116}.

\bibitem{cramer_efficient_2010}
\bibinfo{author}{Cramer, M.} \emph{et~al.}
\newblock \bibinfo{title}{Efficient quantum state tomography}.
\newblock \emph{\bibinfo{journal}{Nature Communications}} \textbf{\bibinfo{volume}{1}}, \bibinfo{pages}{149} (\bibinfo{year}{2010}).
\newblock \urlprefix\url{https://www.nature.com/articles/ncomms1147}.

\bibitem{magesan_scalable_2011}
\bibinfo{author}{Magesan, E.}, \bibinfo{author}{Gambetta, J.~M.} \& \bibinfo{author}{Emerson, J.}
\newblock \bibinfo{title}{Scalable and {Robust} {Randomized} {Benchmarking} of {Quantum} {Processes}}.
\newblock \emph{\bibinfo{journal}{Physical Review Letters}} \textbf{\bibinfo{volume}{106}}, \bibinfo{pages}{180504} (\bibinfo{year}{2011}).
\newblock \urlprefix\url{https://link.aps.org/doi/10.1103/PhysRevLett.106.180504}.
\newblock \bibinfo{note}{Publisher: American Physical Society}.

\bibitem{magesan_efficient_2012}
\bibinfo{author}{Magesan, E.} \emph{et~al.}
\newblock \bibinfo{title}{Efficient {Measurement} of {Quantum} {Gate} {Error} by {Interleaved} {Randomized} {Benchmarking}}.
\newblock \emph{\bibinfo{journal}{Physical Review Letters}} \textbf{\bibinfo{volume}{109}}, \bibinfo{pages}{080505} (\bibinfo{year}{2012}).
\newblock \urlprefix\url{https://link.aps.org/doi/10.1103/PhysRevLett.109.080505}.

\bibitem{gaebler_randomized_2012}
\bibinfo{author}{Gaebler, J.~P.} \emph{et~al.}
\newblock \bibinfo{title}{Randomized {Benchmarking} of {Multiqubit} {Gates}}.
\newblock \emph{\bibinfo{journal}{Physical Review Letters}} \textbf{\bibinfo{volume}{108}}, \bibinfo{pages}{260503} (\bibinfo{year}{2012}).
\newblock \urlprefix\url{https://link.aps.org/doi/10.1103/PhysRevLett.108.260503}.

\bibitem{onorati_randomized_2018}
\bibinfo{author}{Onorati, E.}, \bibinfo{author}{Werner, A.~H.} \& \bibinfo{author}{Eisert, J.}
\newblock \bibinfo{title}{Randomized benchmarking for individual quantum gates}.
\newblock \emph{\bibinfo{journal}{arXiv:1811.11775 [quant-ph]}}  (\bibinfo{year}{2018}).
\newblock \urlprefix\url{http://arxiv.org/abs/1811.11775}.
\newblock \bibinfo{note}{ArXiv: 1811.11775}.

\bibitem{cross_validating_2018}
\bibinfo{author}{Cross, A.~W.}, \bibinfo{author}{Bishop, L.~S.}, \bibinfo{author}{Sheldon, S.}, \bibinfo{author}{Nation, P.~D.} \& \bibinfo{author}{Gambetta, J.~M.}
\newblock \bibinfo{title}{Validating quantum computers using randomized model circuits}.
\newblock \emph{\bibinfo{journal}{arXiv:1811.12926 [quant-ph]}}  (\bibinfo{year}{2018}).
\newblock \urlprefix\url{http://arxiv.org/abs/1811.12926}.
\newblock \bibinfo{note}{ArXiv: 1811.12926}.

\bibitem{nielsen_gate_2021}
\bibinfo{author}{Nielsen, E.} \emph{et~al.}
\newblock \bibinfo{title}{Gate {Set} {Tomography}}.
\newblock \emph{\bibinfo{journal}{Quantum}} \textbf{\bibinfo{volume}{5}}, \bibinfo{pages}{557} (\bibinfo{year}{2021}).
\newblock \urlprefix\url{https://quantum-journal.org/papers/q-2021-10-05-557/}.
\newblock \bibinfo{note}{Publisher: Verein zur Förderung des Open Access Publizierens in den Quantenwissenschaften}.

\bibitem{Poyatos1997}
\bibinfo{author}{Poyatos, J.~F.}, \bibinfo{author}{Cirac, J.~I.} \& \bibinfo{author}{Zoller, P.}
\newblock \bibinfo{title}{{Complete Characterization of a Quantum Process: The Two-Bit Quantum Gate}}.
\newblock \emph{\bibinfo{journal}{Physical Review Letters}} \textbf{\bibinfo{volume}{78}}, \bibinfo{pages}{390--393} (\bibinfo{year}{1997}).

\bibitem{Nielsen00}
\bibinfo{author}{Nielsen, M.~A.} \& \bibinfo{author}{Chuang, I.~L.}
\newblock \emph{\bibinfo{title}{Quantum Computation and Quantum Information}} (\bibinfo{publisher}{Cambridge University Press}, \bibinfo{year}{2000}).

\bibitem{Merkel2013}
\bibinfo{author}{Merkel, S.~T.} \emph{et~al.}
\newblock \bibinfo{title}{{Self-consistent quantum process tomography}}.
\newblock \emph{\bibinfo{journal}{Physical Review A}} \textbf{\bibinfo{volume}{062119}}, \bibinfo{pages}{1--9} (\bibinfo{year}{2013}).

\bibitem{Blume-kohout2017}
\bibinfo{author}{Blume-kohout, R.} \emph{et~al.}
\newblock \bibinfo{title}{{Demonstration of qubit operations below a rigorous fault tolerance threshold with gate set tomography}}.
\newblock \emph{\bibinfo{journal}{Nature Communications}}  (\bibinfo{year}{2017}).

\bibitem{Greenbaum2015}
\bibinfo{author}{Greenbaum, D.}
\newblock \bibinfo{title}{{Introduction to Quantum Gate Set Tomography}}  (\bibinfo{year}{2015}).
\newblock \eprint{arXiv:1509.02921v1}.

\bibitem{gross_quantum_2010}
\bibinfo{author}{Gross, D.}, \bibinfo{author}{Liu, Y.-K.}, \bibinfo{author}{Flammia, S.~T.}, \bibinfo{author}{Becker, S.} \& \bibinfo{author}{Eisert, J.}
\newblock \bibinfo{title}{Quantum {State} {Tomography} via {Compressed} {Sensing}}.
\newblock \emph{\bibinfo{journal}{Physical Review Letters}} \textbf{\bibinfo{volume}{105}}, \bibinfo{pages}{150401} (\bibinfo{year}{2010}).
\newblock \urlprefix\url{http://link.aps.org/doi/10.1103/PhysRevLett.105.150401}.

\bibitem{flammia_quantum_2012}
\bibinfo{author}{Flammia, S.~T.}, \bibinfo{author}{Gross, D.}, \bibinfo{author}{Liu, Y.-K.} \& \bibinfo{author}{Eisert, J.}
\newblock \bibinfo{title}{Quantum tomography via compressed sensing: error bounds, sample complexity and efficient estimators}.
\newblock \emph{\bibinfo{journal}{New Journal of Physics}} \textbf{\bibinfo{volume}{14}}, \bibinfo{pages}{095022} (\bibinfo{year}{2012}).
\newblock \urlprefix\url{http://stacks.iop.org/1367-2630/14/i=9/a=095022}.

\bibitem{PhysRevLett.119.180509}
\bibinfo{author}{Temme, K.}, \bibinfo{author}{Bravyi, S.} \& \bibinfo{author}{Gambetta, J.~M.}
\newblock \bibinfo{title}{Error mitigation for short-depth quantum circuits}.
\newblock \emph{\bibinfo{journal}{Phys. Rev. Lett.}} \textbf{\bibinfo{volume}{119}}, \bibinfo{pages}{180509} (\bibinfo{year}{2017}).
\newblock \urlprefix\url{https://link.aps.org/doi/10.1103/PhysRevLett.119.180509}.

\bibitem{huszar_adaptive_2012}
\bibinfo{author}{Huszár, F.} \& \bibinfo{author}{Houlsby, N.}
\newblock \bibinfo{title}{Adaptive {Bayesian} quantum tomography}.
\newblock \emph{\bibinfo{journal}{Physical Review A}} \textbf{\bibinfo{volume}{85}}, \bibinfo{pages}{052120} (\bibinfo{year}{2012}).
\newblock \urlprefix\url{https://link.aps.org/doi/10.1103/PhysRevA.85.052120}.
\newblock \bibinfo{note}{Publisher: American Physical Society}.

\bibitem{ferrie_self-guided_2014}
\bibinfo{author}{Ferrie, C.}
\newblock \bibinfo{title}{Self-{Guided} {Quantum} {Tomography}}.
\newblock \emph{\bibinfo{journal}{Physical Review Letters}} \textbf{\bibinfo{volume}{113}}, \bibinfo{pages}{190404} (\bibinfo{year}{2014}).
\newblock \urlprefix\url{https://link.aps.org/doi/10.1103/PhysRevLett.113.190404}.

\bibitem{kueng_near-optimal_2015}
\bibinfo{author}{Kueng, R.} \& \bibinfo{author}{Ferrie, C.}
\newblock \bibinfo{title}{Near-optimal quantum tomography: estimators and bounds}.
\newblock \emph{\bibinfo{journal}{New Journal of Physics}} \textbf{\bibinfo{volume}{17}}, \bibinfo{pages}{123013} (\bibinfo{year}{2015}).
\newblock \urlprefix\url{https://doi.org/10.1088%2F1367-2630%2F17%2F12%2F123013}.
\newblock \bibinfo{note}{Publisher: IOP Publishing}.

\bibitem{granade_practical_2016}
\bibinfo{author}{Granade, C.}, \bibinfo{author}{Combes, J.} \& \bibinfo{author}{Cory, D.~G.}
\newblock \bibinfo{title}{Practical {Bayesian} tomography}.
\newblock \emph{\bibinfo{journal}{New Journal of Physics}} \textbf{\bibinfo{volume}{18}}, \bibinfo{pages}{033024} (\bibinfo{year}{2016}).
\newblock \urlprefix\url{https://doi.org/10.1088%2F1367-2630%2F18%2F3%2F033024}.
\newblock \bibinfo{note}{Publisher: IOP Publishing}.

\bibitem{PhysRevA.108.022602}
\bibinfo{author}{Milazzo, N.}, \bibinfo{author}{Giraud, O.}, \bibinfo{author}{Gramegna, G.} \& \bibinfo{author}{Braun, D.}
\newblock \bibinfo{title}{Principles of quantum functional testing}.
\newblock \emph{\bibinfo{journal}{Phys. Rev. A}} \textbf{\bibinfo{volume}{108}}, \bibinfo{pages}{022602} (\bibinfo{year}{2023}).
\newblock \urlprefix\url{https://link.aps.org/doi/10.1103/PhysRevA.108.022602}.

\bibitem{PRXQuantum.2.020303}
\bibinfo{author}{Fiderer, L.~J.}, \bibinfo{author}{Schuff, J.} \& \bibinfo{author}{Braun, D.}
\newblock \bibinfo{title}{Neural-network heuristics for adaptive bayesian quantum estimation}.
\newblock \emph{\bibinfo{journal}{PRX Quantum}} \textbf{\bibinfo{volume}{2}}, \bibinfo{pages}{020303} (\bibinfo{year}{2021}).
\newblock \urlprefix\url{https://link.aps.org/doi/10.1103/PRXQuantum.2.020303}.

\bibitem{quek_adaptive_2021}
\bibinfo{author}{Quek, Y.}, \bibinfo{author}{Fort, S.} \& \bibinfo{author}{Ng, H.~K.}
\newblock \bibinfo{title}{Adaptive quantum state tomography with neural networks}.
\newblock \emph{\bibinfo{journal}{npj Quantum Information}} \textbf{\bibinfo{volume}{7}}, \bibinfo{pages}{1--7} (\bibinfo{year}{2021}).
\newblock \urlprefix\url{https://www.nature.com/articles/s41534-021-00436-9}.
\newblock \bibinfo{note}{Number: 1 Publisher: Nature Publishing Group}.

\bibitem{gebhart_learning_2023}
\bibinfo{author}{Gebhart, V.} \emph{et~al.}
\newblock \bibinfo{title}{Learning quantum systems}.
\newblock \emph{\bibinfo{journal}{Nature Reviews Physics}} \textbf{\bibinfo{volume}{5}}, \bibinfo{pages}{141--156} (\bibinfo{year}{2023}).
\newblock \urlprefix\url{https://www.nature.com/articles/s42254-022-00552-1}.
\newblock \bibinfo{note}{Number: 3 Publisher: Nature Publishing Group}.

\bibitem{huang_predicting_2020}
\bibinfo{author}{Huang, H.-Y.}, \bibinfo{author}{Kueng, R.} \& \bibinfo{author}{Preskill, J.}
\newblock \bibinfo{title}{Predicting many properties of a quantum system from very few measurements}.
\newblock \emph{\bibinfo{journal}{Nature Physics}} \textbf{\bibinfo{volume}{16}}, \bibinfo{pages}{1050--1057} (\bibinfo{year}{2020}).
\newblock \urlprefix\url{https://www.nature.com/articles/s41567-020-0932-7}.
\newblock \bibinfo{note}{Number: 10 Publisher: Nature Publishing Group}.

\bibitem{Mil.Bra.Gir:19}
\bibinfo{author}{Milazzo, N.}, \bibinfo{author}{Braun, D.} \& \bibinfo{author}{Giraud, O.}
\newblock \bibinfo{title}{Optimal measurement strategies for fast entanglement detection}.
\newblock \emph{\bibinfo{journal}{Phys. Rev. A}} \textbf{\bibinfo{volume}{100}}, \bibinfo{pages}{012328} (\bibinfo{year}{2019}).

\bibitem{PhysRevA.96.032312}
\bibinfo{author}{Bohnet-Waldraff, F.}, \bibinfo{author}{Braun, D.} \& \bibinfo{author}{Giraud, O.}
\newblock \bibinfo{title}{Entanglement and the truncated moment problem}.
\newblock \emph{\bibinfo{journal}{Phys. Rev. A}} \textbf{\bibinfo{volume}{96}}, \bibinfo{pages}{032312} (\bibinfo{year}{2017}).
\newblock \urlprefix\url{https://link.aps.org/doi/10.1103/PhysRevA.96.032312}.

\bibitem{PhysRevA.102.052406}
\bibinfo{author}{Milazzo, N.}, \bibinfo{author}{Braun, D.} \& \bibinfo{author}{Giraud, O.}
\newblock \bibinfo{title}{Truncated moment sequences and a solution to the channel separability problem}.
\newblock \emph{\bibinfo{journal}{Phys. Rev. A}} \textbf{\bibinfo{volume}{102}}, \bibinfo{pages}{052406} (\bibinfo{year}{2020}).
\newblock \urlprefix\url{https://link.aps.org/doi/10.1103/PhysRevA.102.052406}.

\bibitem{gupta_semiconductor_2009}
\bibinfo{author}{Gupta, A.}
\newblock \emph{\bibinfo{title}{Semiconductor {Memory} {Testing}: {Fault} {Models} and {Test} {Considerations} for {High} {Performance} {Embedded} {SRAM}'s}} (\bibinfo{publisher}{VDM Verlag}, \bibinfo{year}{2009}).

\bibitem{Wang2014}
\bibinfo{author}{Wang, C.} \emph{et~al.}
\newblock \bibinfo{title}{{Measurement and control of quasiparticle dynamics in a superconducting qubit}}.
\newblock \emph{\bibinfo{journal}{Nature Communications}} \textbf{\bibinfo{volume}{5}}, \bibinfo{pages}{1--7} (\bibinfo{year}{2014}).
\newblock \eprint{1406.7300}.

\bibitem{Pan2022}
\bibinfo{author}{Pan, X.} \emph{et~al.}
\newblock \bibinfo{title}{{Engineering superconducting qubits to reduce quasiparticles and charge noise}}.
\newblock \emph{\bibinfo{journal}{Nature Communications}} \textbf{\bibinfo{volume}{13}}, \bibinfo{pages}{1--7} (\bibinfo{year}{2022}).
\newblock \eprint{2202.01435}.

\bibitem{Hahn1950}
\bibinfo{author}{Hahn, E.~L.}
\newblock \bibinfo{title}{{Spin echoes}}.
\newblock \emph{\bibinfo{journal}{Physical Review}} \textbf{\bibinfo{volume}{80}}, \bibinfo{pages}{580--594} (\bibinfo{year}{1950}).

\bibitem{Burkard2023}
\bibinfo{author}{Gul{\'{a}}csi, B.} \& \bibinfo{author}{Burkard, G.}
\newblock \bibinfo{title}{{Signatures of non-Markovianity of a superconducting qubit}}.
\newblock \emph{\bibinfo{journal}{Physical Review B}} \textbf{\bibinfo{volume}{107}} (\bibinfo{year}{2023}).
\newblock \eprint{2302.09092v1}.

\bibitem{Agarwal2023}
\bibinfo{author}{Agarwal, A.}, \bibinfo{author}{Lindoy, L.~P.}, \bibinfo{author}{Lall, D.}, \bibinfo{author}{Jamet, F.} \& \bibinfo{author}{Rungger, I.}
\newblock \bibinfo{title}{{Modelling non-Markovian noise in driven superconducting qubits}} \bibinfo{pages}{1--26} (\bibinfo{year}{2023}).
\newblock \eprint{2306.13021}.

\bibitem{Lisenfeld2019}
\bibinfo{author}{Lisenfeld, J.} \emph{et~al.}
\newblock \bibinfo{title}{{Electric field spectroscopy of material defects in transmon qubits}}.
\newblock \emph{\bibinfo{journal}{npj Quantum Information}} \bibinfo{pages}{1--6} (\bibinfo{year}{2019}).

\bibitem{Mueller2015}
\bibinfo{author}{M\"uller, C.}, \bibinfo{author}{Lisenfeld, J.}, \bibinfo{author}{Shnirman, A.} \& \bibinfo{author}{Poletto, S.}
\newblock \bibinfo{title}{{Interacting two-level defects as sources of fluctuating high-frequency noise in superconducting circuits}}.
\newblock \emph{\bibinfo{journal}{Physical Review B}} \textbf{\bibinfo{volume}{035442}}, \bibinfo{pages}{1--13} (\bibinfo{year}{2015}).

\bibitem{Mueller2019}
\bibinfo{author}{M\"uller, C.}, \bibinfo{author}{Cole, J.} \& \bibinfo{author}{Lisenfeld, J.}
\newblock \bibinfo{title}{{Towards understanding two-level-systems in amorphous solids: insights from quantum circuits}}.
\newblock \emph{\bibinfo{journal}{Reports on Progress in Physics}} \textbf{\bibinfo{volume}{82}} (\bibinfo{year}{2019}).

\bibitem{Perrin2024}
\bibinfo{author}{Perrin, H.}, \bibinfo{author}{Scoquart, T.}, \bibinfo{author}{Shnirman, A.}, \bibinfo{author}{Schmalian, J.} \& \bibinfo{author}{Snizhko, K.}
\newblock \bibinfo{title}{{Mitigating crosstalk errors by randomized compiling: Simulation of the BCS model on a superconducting quantum computer}}.
\newblock \emph{\bibinfo{journal}{Physical Review Letters}} \textbf{\bibinfo{volume}{013142}}, \bibinfo{pages}{1--24} (\bibinfo{year}{2024}).

\bibitem{Gambetta2012}
\bibinfo{author}{Gambetta, J.~M.} \emph{et~al.}
\newblock \bibinfo{title}{{Characterization of Addressability by Simultaneous Randomized Benchmarking}}.
\newblock \emph{\bibinfo{journal}{Physical Review Letters}} \textbf{\bibinfo{volume}{240504}}, \bibinfo{pages}{1--5} (\bibinfo{year}{2012}).

\bibitem{Heunisch2023}
\bibinfo{author}{Heunisch, L.}, \bibinfo{author}{Eichler, C.} \& \bibinfo{author}{Hartmann, M.~J.}
\newblock \bibinfo{title}{{Tunable coupler to fully decouple and maximally localize superconducting qubits}}.
\newblock \emph{\bibinfo{journal}{Physical Review Applied}} \textbf{\bibinfo{volume}{10}}, \bibinfo{pages}{1} (\bibinfo{year}{2023}).
\newblock \urlprefix\url{https://doi.org/10.1103/PhysRevApplied.20.064037}.

\bibitem{Sung2021}
\bibinfo{author}{Sung, Y.} \emph{et~al.}
\newblock \bibinfo{title}{{Realization of High-Fidelity CZ and ZZ -Free iSWAP Gates with a Tunable Coupler}}.
\newblock \emph{\bibinfo{journal}{Physical Review X}} \textbf{\bibinfo{volume}{11}}, \bibinfo{pages}{21058} (\bibinfo{year}{2021}).
\newblock \urlprefix\url{https://doi.org/10.1103/PhysRevX.11.021058}.

\bibitem{Murali2020}
\bibinfo{author}{Murali, P.}, \bibinfo{author}{McKay, D.~C.}, \bibinfo{author}{Martonosi, M.} \& \bibinfo{author}{Javid-Abhari, A.}
\newblock \bibinfo{title}{{Software Mitigation of Crosstalk on Noisy Intermediate-Scale Quantum Computers}}.
\newblock \emph{\bibinfo{journal}{ASPLOS '20: Proceedings of the Twenty-Fifth International Conference on Architectural Support for Programming Languages and Operating Systems}} \bibinfo{pages}{1001--1016} (\bibinfo{year}{2020}).

\bibitem{Samach2022}
\bibinfo{author}{Samach, G.~O.} \emph{et~al.}
\newblock \bibinfo{title}{{Lindblad Tomography of a Superconducting Quantum Processor}}.
\newblock \emph{\bibinfo{journal}{Physical Review Applied}} \textbf{\bibinfo{volume}{18}}, \bibinfo{pages}{1} (\bibinfo{year}{2022}).
\newblock \eprint{2105.02338}.

\bibitem{OMalley2015}
\bibinfo{author}{O'Malley, P.~J.} \emph{et~al.}
\newblock \bibinfo{title}{{Qubit Metrology of Ultralow Phase Noise Using Randomized Benchmarking}}.
\newblock \emph{\bibinfo{journal}{Physical Review Applied}} \textbf{\bibinfo{volume}{3}}, \bibinfo{pages}{1--11} (\bibinfo{year}{2015}).
\newblock \eprint{1411.2613}.

\bibitem{Jozsa03}
\bibinfo{author}{Jozsa, R.} \& \bibinfo{author}{Linden, N.}
\newblock \bibinfo{title}{On the role of entanglement in quantum computational speed-up}.
\newblock \emph{\bibinfo{journal}{Proc. R. Soc. Lond. A}} \textbf{\bibinfo{volume}{459}}, \bibinfo{pages}{2011--2032} (\bibinfo{year}{2003}).

\bibitem{Ketterer2023}
\bibinfo{author}{Ketterer, A.} \& \bibinfo{author}{Wellens, T.}
\newblock \bibinfo{title}{{Characterizing crosstalk of superconducting transmon processors}}.
\newblock \emph{\bibinfo{journal}{Physical Review Applied}} \textbf{\bibinfo{volume}{10}}, \bibinfo{pages}{1} (\bibinfo{year}{2023}).
\newblock \eprint{2303.14103}.

\bibitem{Hertzberg2021}
\bibinfo{author}{Hertzberg, J.~B.} \emph{et~al.}
\newblock \bibinfo{title}{Laser-annealing josephson junctions for yielding scaled-up superconducting quantum processors}.
\newblock \emph{\bibinfo{journal}{npj Quantum Information}} \textbf{\bibinfo{volume}{7}} (\bibinfo{year}{2021}).

\bibitem{Krantz2019}
\bibinfo{author}{Krantz, P.} \emph{et~al.}
\newblock \bibinfo{title}{{A quantum engineer's guide to superconducting qubits}}.
\newblock \emph{\bibinfo{journal}{Applied Physics Reviews}} \textbf{\bibinfo{volume}{6}} (\bibinfo{year}{2019}).
\newblock \eprint{1904.06560}.

\bibitem{IBMQuantum}
\bibinfo{title}{{IBM Quantum Platform}} (\bibinfo{year}{2024}).
\newblock \urlprefix\url{https://quantum.ibm.com/}.

\bibitem{Johansson2013}
\bibinfo{author}{Johansson, J.~R.}, \bibinfo{author}{Nation, P.~D.} \& \bibinfo{author}{Nori, F.}
\newblock \bibinfo{title}{{QuTiP 2: A Python framework for the dynamics of open quantum systems}}.
\newblock \emph{\bibinfo{journal}{Computer Physics Communications}} \textbf{\bibinfo{volume}{184}}, \bibinfo{pages}{1234--1240} (\bibinfo{year}{2013}).
\newblock \eprint{1211.6518}.

\end{thebibliography}

\onecolumngrid
\newpage

\setcounter{equation}{0}
\setcounter{page}{1}
\setcounter{figure}{0}
\setcounter{table}{0}
\setcounter{section}{0}

\makeatletter
\renewcommand{\theequation}{S\arabic{equation}}
\renewcommand{\thefigure}{S\arabic{figure}}
\renewcommand{\thetable}{S\arabic{table}}

\begin{center}
{\Large SUPPLEMENTAL MATERIAL}
\end{center}
\begin{center}
\vspace{0.8cm}
{\Large Pattern-based quantum functional testing}
\end{center}
\begin{center}
Erik Weiss,$^{1}$ Marcel Cech,$^{1}$ Stanislaw Soltan,$^{1}$ Martin Koppenhöfer,$^{2}$ \\ Michael Krebsbach,$^{2}$ Thomas Wellens ,$^{2}$ and Daniel Braun,$^{1}$
\end{center}
\begin{center}
$^1${\em Institut f\"ur theoretische Physik, Universit\"at T\"{u}bingen, 72076 T\"ubingen, Germany}\\
$^2${\em Fraunhofer Institute for Applied Solid State Physics IAF, Tullastr. 72, Freiburg 79108, Germany}\\
\end{center}

\section{Circuits}\label{sec:Circuits}

\begin{figure}
	\includegraphics[width=\linewidth]{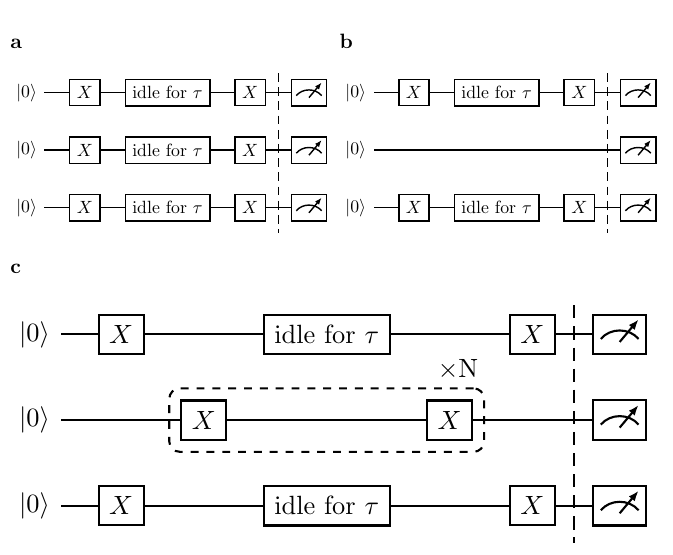}
	\caption{\textbf{Circuits of $\ket{1}$ patterns}.
	\textbf{a} Circuit for a blank $\ket{1}$ pattern. \textbf{b} Circuit for a checkerboard $\ket{1}$ pattern.
	\textbf{c} Circuit for a checkerboard $\ket{1}$ pattern with active spectator qubits. Instead of leaving spectator qubits idle in $\ket{0}$ we apply an even number $N$ of X-gates to spectator qubits.
	}
	\label{fig:S1}
\end{figure}

\begin{figure}
	\includegraphics[width=\linewidth]{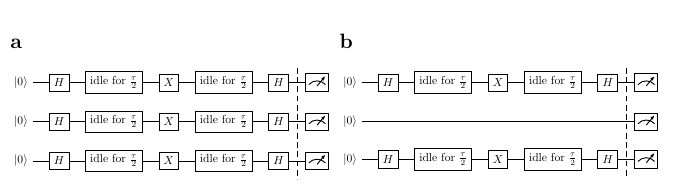}
	\caption{\textbf{Circuits for single qubit superposition patterns}.
	\textbf{a} Circuit for an echoed blank $\ket{+}$ pattern. \textbf{b} Circuit for an echoed checkerboard $\ket{+}$ pattern. 
	}
	\label{fig:S2}
\end{figure}

\begin{figure}
	\includegraphics[width=\linewidth]{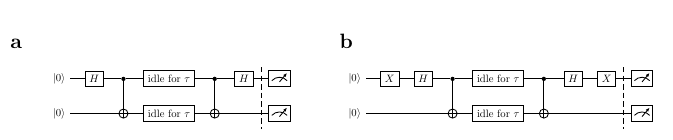}
	\caption{\textbf{Circuits to test the lifetime of Bell states}.
	After preparing a Bell state, qubits are left idle for a duration $\tau$ and then the inverse gate sequence is applied to disentangle them and bring them back to their state $\ket{0}$.
	\textbf{a} Circuit to test the lifetime of a $\ket{\Phi^+}$ state. \textbf{b} Circuit to test the lifetime of a $\ket{\Phi^-}$ state.
	}
	\label{fig:S3}
\end{figure}

Here, we present the circuits not explicitly shown in the main document. For simplicity, we provide an example with three qubits. At the end of each circuit, all physical qubits are measured. Figure~\ref{fig:S1} illustrates the circuits used for $\ket{1}$ patterns and Figure~\ref{fig:S2} shows circuits for $\ket{+}$ patterns. To investigate how long a two-qubit entangled state, like the $\ket{\Phi^+}$ or $\ket{\Phi^-}$ Bell-states, can be maintained within a quantum device, we use circuits such as the ones depicted in Figure~\ref{fig:S3}.

\section{Hardware}\label{sec:Hardware}
\begin{figure}
	\includegraphics[width=\linewidth]{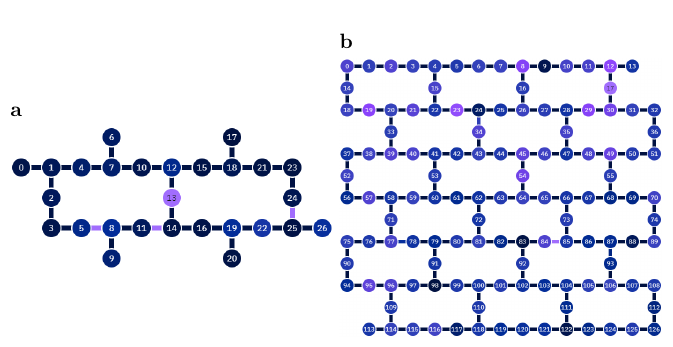}
	\caption{\textbf{Chip architectures of two different quantum processors}.
	Nodes represent qubits and edges represent physical couplings between them. \textbf{a} \textit{ibmq\_ehningen} based on 27 qubit Falcon processor \textbf{b} \textit{ibm\_brisbane} based on 127 qubit Eagle processor\cite{IBMQuantum}.
	}
	\label{fig:S4}
\end{figure}

To demonstrate the effectiveness of our pattern-based testing approach, we tested our method on quantum computers with superconducting fixed frequency transmon qubits developed by IBM. two types of devices were used, the 27-qubit device named \textit{ibmq\_ehningen}, based on the Falcon r5.11 architecture, and the 127-qubit device \textit{ibm\_brisbane}, based on the Eagle r3 architecture. Qubit layouts of both devices are shown in Figure~\ref{fig:S4}.

\begin{figure}
	\includegraphics[width=\linewidth]{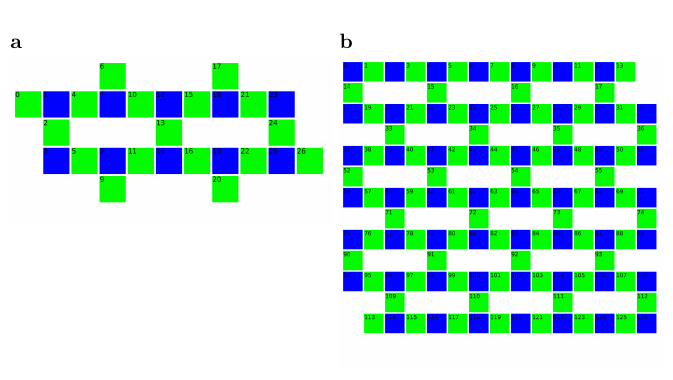}
	\caption{\textbf{Checkerboard patterns}.
	We define checkerboard pattern A as the configuration where green qubits are the target qubits and blue qubits are the spectators. Conversely, in checkerboard pattern B, the blue qubits are the target qubits and the green qubits are the spectators.
	\textbf{a} \textit{ibmq\_ehningen} \textbf{b} \textit{ibm\_brisbane}. 
	}
	\label{fig:S5}
\end{figure}

For a clear definition of which qubits we refer to as target or spectator qubits in various checkerboard patterns we define them in Figure~\ref{fig:S5}.

\section{Simulation of $\ket{1}$ pattern and $\ket{+}$ pattern with interacting qubits}\label{sec:Simulations of patterns}

To explain why there are fidelity oscillations in blank $\ket{+}$ patterns but not in blank $\ket{1}$ patterns, we simulate different $\ket{1}$ and $\ket{+}$ patterns for the two-qubit Hamiltonian $H = \Omega_{\alpha \alpha} \sigma_{\alpha} \otimes \sigma_{\alpha}$ with $\alpha \in \{x,y,z\}$. We then compare these simulations with experimental data of Qubit 20 of \textit{ibmq\_ehningen}, which has only one nearest neighbor (qubit 19).

For the simulations, we utilize the Lindblad Master Equation Solver available in the python library QuTiP \cite{Johansson2013}. It employs the following form of the Master equation:
\begin{align}
	\dot{\rho}(t) = -\frac{i}{\hbar} [H,\rho(t)] +\sum_{n}\frac{1}{2}(2C_n\rho(t)C_n^\dagger -\rho(t)C_n^\dagger C_n - C_n^\dagger C_n\rho(t)).
\end{align}
With the transition operators $C_n=\sqrt{\gamma}A_n$, where $A_n$ are the system coupling agents through which the environment couples to the system and $\gamma_n$ the corresponding rates.
In case of $\ket{1}$ patterns, we use  $C_0 = \frac{1}{\sqrt{T_1}}(\sigma_+ \otimes \mathbb{1})$ and $C_1 = \frac{1}{\sqrt{T_1}}(\mathbb{1} \otimes\sigma_+)$ with 
$\sigma_+ = \ket{1}\bra{0}$. In case of $\ket{+}$ patterns, $C_0 = \frac{1}{\sqrt{T_2}}(\sigma_z \otimes \mathbb{1})$ and $C_1 = \frac{1}{\sqrt{T_2}}(\mathbb{1} \otimes\sigma_z)$.

\begin{figure}
	\includegraphics[width=\linewidth]{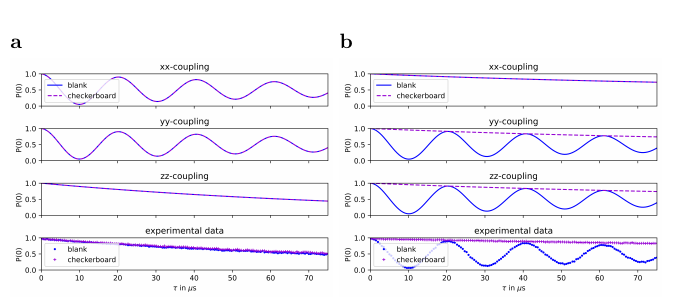}
	\caption{\textbf{Simulations of $\ket{1}$ and $\ket{+}$ patterns on two coupled qubits}.
	Qubit dynamics are simulated with a coupling term of the form $H = \Omega_{\alpha \alpha} \sigma_{\alpha} \otimes \sigma_{\alpha}$ with $\alpha \in \{x,y,z\}$. Only the dynamics of one of the two qubits is displayed. A checkerboard pattern in this setup means, that one of the two coupled qubits remains in $\ket{0}$. Simulations are compared with experimental data from qubit 20 of \textit{ibmq\_ehningen}.
	\textbf{a} $\ket{1}$ patterns \textbf{b} $\ket{+}$ patterns. For our simulations, we utilize the Lindblad Master Equation Solver available in the python library \cite{Johansson2013}. In case of $\ket{1}$ patterns we use the transition operators  $C_0 = \frac{1}{\sqrt{T_1}}(\sigma_+ \otimes \mathbb{1})$ and $C_1 = \frac{1}{\sqrt{T_1}}(\mathbb{1} \otimes\sigma_+)$. For $\ket{+}$ patterns the transition operators are $C_0 = \frac{1}{\sqrt{T_2}}(\sigma_z \otimes \mathbb{1})$ and $C_1 = \frac{1}{\sqrt{T_2}}(\mathbb{1} \otimes\sigma_z)$. The simulation parameters used are  $\Omega_{\alpha \alpha}/2\pi = 1.54 \,\mathrm{MHz}$, and for simplicity we set $T_1= 92 \,\mathrm{\mu s}$ and $T_2= 204  \,\mathrm{\mu s}$ for both qubits.
	Only the $zz$-coupling is able to reproduce the experimental data for all patterns.
	}
	\label{fig:S6}
\end{figure}

From Figure~\ref{fig:S6}, it becomes clear that of the different couplings tested only $zz$-coupling is capable of explaining both data from $\ket{1}$ and $\ket{+}$ patterns.

\section{Size limit of entangled states}\label{sec:Size limit of entangled states}

To assess the viability of investigating patterns containing entangled states with many qubits, we examined the fidelity of entangled states composed of $N$ qubits. Specifically we used states of the form
\begin{align}
	\ket{\text{GHZ}} = \frac{\ket{0}^{\otimes N} + \ket{1}^{\otimes N}}{\sqrt{2}}.
\end{align}
Given the limited connectivity in current quantum devices, we randomly selected $n=10$ connected chains of qubits for various sizes $N$ such that the corresponding GHZ states can be generated by applying $N-1$ subsequent CNOT gates without additional SWAP gates. Similar to the circuit depicted in Figure~\ref{fig:S3} for the $N=2$ GHZ state, we assess the lifetime of such states by first preparing the GHZ state then wait for a time $\tau$ after which we undo all operations and measure all involved qubits in the computational basis. To decrease sensitivity to quasistatic, low-frequency noise, we place an X-gate at $\tau/2$.  We then calculate the fidelity  $F(\ket{0}^{\otimes N}\bra{0}^{\otimes N}, \rho_{\text{pi}})$ and average it over all $n=10$ samples.

\begin{figure}
	\includegraphics[width=\linewidth]{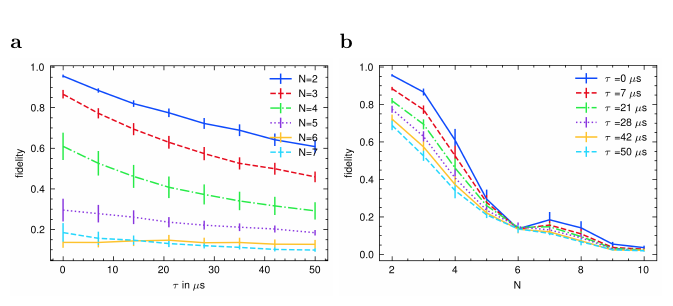}
	\caption{\textbf{Fidelity of GHZ states with different sizes $N$}.
	\textbf{a} Mean fidelity as a function of delay time $\tau$. \textbf{b} Mean fidelity as a function of the size $N$ of a GHZ state for different delay times $\tau$. Error bars give the estimator of the standard error $\sigma_{\bar{F}} = \sigma_F / \sqrt{n}$.
	}
	\label{fig:S7}
\end{figure}

As becomes clear from Figure~\ref{fig:S7} achieving high-fidelity entanglement with a significant number of qubits remains challenging in existing devices.

\end{document}